\documentclass[aps,nofootinbib,amsmath,prd,onecolumn,notitlepage,showpacs,superscriptaddress,11pt]{revtex4-1}

\usepackage[american]{babel}
\usepackage{amsfonts}
\usepackage{amsmath}
\usepackage{amssymb}
\usepackage{slashed}
\usepackage{upgreek}

\usepackage{xcolor}
\usepackage{bm}
\usepackage{float}
\usepackage[utf8]{inputenc}
\usepackage{nicefrac}
\usepackage{hyperref}
\usepackage{url}
\usepackage{tikz-cd}
\usepackage[normalem]{ulem}
\usepackage{mathtools}
\usepackage{graphicx,txfonts}

\definecolor{darkblue}{rgb}{0.1,0.1,0.7}
\hypersetup{colorlinks,
           linkcolor={darkblue},
           citecolor={darkblue},
           urlcolor={darkblue}
}

\newcommand{\lvec}[2]{\raise #1\hbox{$^\leftarrow$} \hspace{-9pt} #2}
\newcommand{\rvec}[2]{\raise #1\hbox{$^\rightarrow$} \hspace{-9pt} #2}
\newcommand{\lrvec}[2]{\raise #1\hbox{$^\leftrightarrow$} \hspace{-9pt} #2}

\allowdisplaybreaks

\usepackage[normalem]{ulem}

\allowdisplaybreaks

\DeclarePairedDelimiter\abs{\lvert}{\rvert}%
\DeclarePairedDelimiter\norm{\lVert}{\rVert}%

\makeatletter
\let\oldabs\abs
\def\abs{\@ifstar{\oldabs}{\oldabs*}}
\let\oldnorm\norm
\def\norm{\@ifstar{\oldnorm}{\oldnorm*}}
\makeatother

\begin{document}

\title{Non-Hermiticity enhanced topological immunity of one-dimensional $p$-wave superconducting chain}

\author{Min Liu}
\thanks{These authors contributed equally to this work.}
\affiliation{Ministry of Education Key Laboratory for Nonequilibrium Synthesis and Modulation of Condensed Matter,Shaanxi Province Key Laboratory of Quantum Information and Quantum Optoelectronic Devices, School of Physics, Xi'an Jiaotong University, Xi'an 710049, China}
\author{Yue Zhang}
\thanks{These authors contributed equally to this work.}
\affiliation{Ministry of Education Key Laboratory for Nonequilibrium Synthesis and Modulation of Condensed Matter,Shaanxi Province Key Laboratory of Quantum Information and Quantum Optoelectronic Devices, School of Physics, Xi'an Jiaotong University, Xi'an 710049, China}
\author{Rui Tian}
\affiliation{Ministry of Education Key Laboratory for Nonequilibrium Synthesis and Modulation of Condensed Matter,Shaanxi Province Key Laboratory of Quantum Information and Quantum Optoelectronic Devices, School of Physics, Xi'an Jiaotong University, Xi'an 710049, China}
\author{Xiayao He}
\affiliation{Ministry of Education Key Laboratory for Nonequilibrium Synthesis and Modulation of Condensed Matter,Shaanxi Province Key Laboratory of Quantum Information and Quantum Optoelectronic Devices, School of Physics, Xi'an Jiaotong University, Xi'an 710049, China}
\author{Tianhao Wu}
\affiliation{Ministry of Education Key Laboratory for Nonequilibrium Synthesis and Modulation of Condensed Matter,Shaanxi Province Key Laboratory of Quantum Information and Quantum Optoelectronic Devices, School of Physics, Xi'an Jiaotong University, Xi'an 710049, China}
\author{Maksims Arzamasovs}
\affiliation{Ministry of Education Key Laboratory for Nonequilibrium Synthesis and Modulation of Condensed Matter,Shaanxi Province Key Laboratory of Quantum Information and Quantum Optoelectronic Devices, School of Physics, Xi'an Jiaotong University, Xi'an 710049, China}

\author{Shuai Li}
\email{lishuai0999@xjtu.edu.cn}
\affiliation{Ministry of Education Key Laboratory for Nonequilibrium Synthesis and Modulation of Condensed Matter,Shaanxi Province Key Laboratory of Quantum Information and Quantum Optoelectronic Devices, School of Physics, Xi'an Jiaotong University, Xi'an 710049, China}
\author{Bo Liu}
\email{liubophy@gmail.com}
\affiliation{Ministry of Education Key Laboratory for Nonequilibrium Synthesis and Modulation of Condensed Matter,Shaanxi Province Key Laboratory of Quantum Information and Quantum Optoelectronic Devices, School of Physics, Xi'an Jiaotong University, Xi'an 710049, China}
\begin{abstract}
Studying the immunity of topological superconductors against non-local disorder is one of the key issues in both fundamental researches and potential applications. Here, we demonstrate that the non-Hermiticity can enhance the robustness of topological edge states against non-local disorder. To illustrate that, we consider a one-dimensional (1D) generalized Kitaev model with the asymmetric hopping in the presence of disorder. It is shown that the region supporting Majorana zero modes (MZMs) against non-local disorder will be enlarged by the non-Hermiticity. Through both the numerical and analytical analyses, we show that non-Hermiticity can stabilize the topological superconducting (SC) phase against higher disorder strength. Our studies would offer new insights into the interplay between non-Hermiticity and topology.
\end{abstract}

\maketitle

\section{\label{sec1}Introduction}
In recent years, the study of topological superconductors (TSC) has attracted considerable attentions in both experimental and theoretical perspectives in various systems \cite{hasan2010colloquium,qi2011topological,armitage2018weyl}. One fantastic property of TSCs is the presence of topological edge states \cite{kane2005quantum,fu2007topological}. One particular example is the Majorana zero modes (MZMs) \cite{majorana1937teoria,read2000paired,nayak2008non,ivanov2001non}, which would be useful in the potential applications in fault-tolerant topological quantum computing \cite{fu2008superconducting,zhang2013majorana,elliott2015colloquium,sun2016majorana,devices1995signatures,wilczek2009majorana}. The Kitaev model, a one-dimensional (1D) spinless $p$-wave superconductor, represents one of the most fundamental and widely studied models for exploring MZMs \cite{kitaev2001unpaired}. Previous studies have shown that the non-local disorder can significantly affect the stability of MZMs, such as leading to their degradation or complete destruction \cite{prodan2010entanglement,altland2010condensed,cai2013topological,pekerten2017disorder,zhang2022topological,tang2020topological,agarwal2023first,ling2024disorder}. Given the practical importance of MZMs, enhancing the robustness of MZMs has emerged as an important problem. In this work, we unveil that the non-Hermiticity can enhance the stability of MZMs against non-local disorder in the Kitaev model.

Such an idea is inspired by recent developments in non-Hermitian topological physics \cite{li2020topological,zhang2023bulk,yao2018non,yokomizo2019non}. As we known, the non-Hermitian topology \cite{yao2018edge,hatano1996localization,bender2007making,moiseyev2011non,el2018non} have created new opportunities for understanding and studying new topological phases. Non-Hermitian systems naturally emerge in open quantum systems, where gain and loss \cite{shen2018quantum,padhan2024complete,kozii2024non} or interactions with the environment \cite{bender1998real,bender2002complex,li2023enhancement} are present. These systems exhibit a variety of novel phenomena, including the non-Hermitian skin effect \cite{lee2016anomalous,yao2018edge,lee2019hybrid,zhang2021acoustic}, exceptional points \cite{choi2010quasieigenstate,hodaei2017enhanced,wang2023experimental}, and the revised bulk-boundary correspondence \cite{yao2018non,yao2018edge,yokomizo2019non,wang2019non}. In the context of TSCs, introducing non-Hermiticity leads to a range of exotic phenomena, such as non-Hermitian MZMs \cite{lieu2019non,mcdonald2018phase,kawabata2018parity,okuma2019topological,li2024anomalous}, complex energy spectra \cite{kawabata2019symmetry,wang2021majorana,yokomizo2021non}, and unconventional quantum phase transitions \cite{cai2021localization, hou2021two}.

In this work, we consider a generalized 1D Kitaev model, where non-local disorder is represented by an incommensurate potential. The non-Hermiticity is captured by the presence of the asymmetric hopping. We conduct both a numerical and analytical investigation of the effects of non-Hermiticity on the robustness of MZMs against non-local disorder. Our results demonstrate that non-Hermiticity significantly enhances the stability of MZMs and the immunity of the topological superconducting (SC) phase against non-local disorder.

\section{\label{sec2}Non-Hermiticity enhance the immunity of TSC}
We consider a 1D non-Hermitian Kitaev chain in the presence of the incommensurate disorder, which can be described by the following Hamiltonian:
\begin{equation}
\mathbf{H}=\sum\limits_{i=1}^{L}[-(t+\gamma /2)\hat{c}_{i}^{\dagger }\hat{c}%
_{i+1}-(t-\gamma /2)\hat{c}_{i+1}^{\dagger }\hat{c}_{i}+\Delta (\hat{c}_{i}%
\hat{c}_{i+1}+h.c.)+V_{i}\hat{n}_{i}],  \label{Eq.Hamiltonian}
\end{equation}
where $L$ is the number of lattice sites. $\hat{c}_{i}^{\dagger }(\hat{c}_{i})$ is the fermion creation (annihilation) operator at lattice site $i$. $\hat{n}_{i}=\hat{c}_{i}^{\dagger }\hat{c}_{i}$ is the particle number operator. $t$ is the symmetric part of the hopping amplitude between the nearest neighbors, while the asymmetric part is denoted by $\gamma $. The $p$-wave pairing amplitude $\Delta$ is taken to be a real number. The incommensurate disorder is captured by $V_{i}=V\cos (2\pi \alpha i )$, with $V$ is the strength of the incommensurate potential. $\alpha$ is an irrational number. The model is reduced to the standard Kitaev $p$-wave model when $\gamma =0$ and $\alpha =0$ \cite{kitaev2001unpaired}.
\begin{figure}[tbp]	
		\centering
         \includegraphics[width=0.6\textwidth]{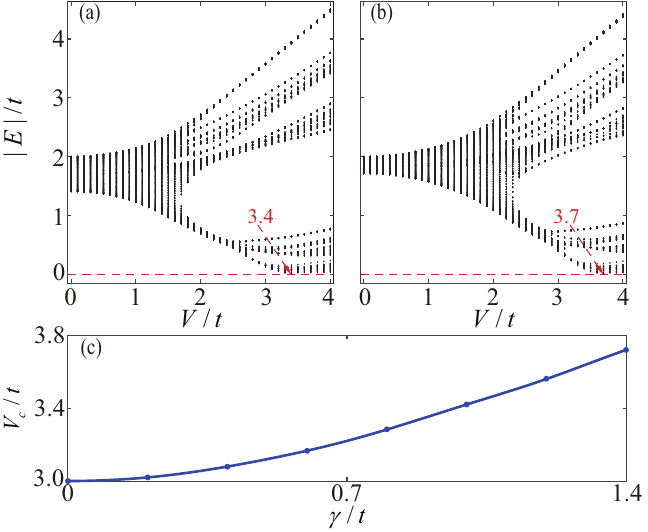}	
		\caption{Absolute values of the eigenenergies of Hamiltonian (\ref{Eq.Hamiltonian}) as a function of $V$ under PBC for (a)$\gamma/t =1$ and (b)$\gamma/t=1.4$. As the disorder strength increases, the energy gap gradually closes. The energy spectrum is divided into two phase regions, separated by the gap-closing point at $V_c$. (c) $V_{c}$ as the function of $\gamma$. It shows a clear increase in $V_c$ with respect to $\gamma$. Here, we choose $\alpha =(\sqrt{5}-1)/2$, $\Delta/t =0.5$ and $L=987$ (a Fibonacci number $F_{m}$).}
		\label{figure:1}		
	\end{figure}
By introducing the Bogoliubov–de Gennes (BdG) transformation \cite{lieb1961two,de2018superconductivity}, $\eta _{n}^{\dagger }=\sum\limits_{i=1}^{L}[u_{n,i}\hat{c}_{i}^{\dagger}+v_{n,i}\hat{c}_{i}]$, the diagonalized Hamiltonian is written as $\mathbf{H}=\sum_{n=1}^{L}E_{n}(\eta_{n}^{\dagger }\eta _{n}-\frac{1}{2})$, where $u_{n}=(u_{n,1},...,u_{n,L})^{T}$ and $v_{n}=(v_{n,1},...,v_{n,L})^{T}$, $E_{n}$ are the $n$-th eigenenergy of the quasiparticles. The energy spectrum and the corresponding eigenwavefunctions can be obtained by diagonalizing the following BdG equation,
\begin{equation}
\begin{pmatrix}
\mathbf{h} & \mathbf{\Delta} \\
-\mathbf{\Delta} & -\mathbf{h}^{\dagger }%
\end{pmatrix}%
\begin{pmatrix}
u_{n} \\
v_{n}%
\end{pmatrix}%
=E_{n}%
\begin{pmatrix}
u_{n} \\
v_{n}%
\end{pmatrix},
\label{Eq.Bdgmatrix}
\end{equation}
where $\mathbf{h}_{i,j}=-(t+\gamma /2)\delta _{j,i+1}-(t-\gamma /2)\delta
_{j,i-1}+V_{i}\delta _{j,i}$, $\mathbf{\Delta}_{i,j}=-\Delta (\delta _{j,i+1}-\delta _{j,i-1})$. When imposing the periodic boundary condition (PBC), through diagonalizing Eq. (\ref{Eq.Bdgmatrix}), we can obtain the eigenspectrum under PBC. Here, the irrational number $\alpha$ is chosen as the inverse of the golden mean, $\alpha = (\sqrt{5}-1)/2$, which is a typical choice in quasiperiodic systems to ensure incommensurability and avoid simple periodicity \cite{cai2013topological,cai2021localization}. It can be approximated by the Fibonacci numbers $F_{m}$ via $\alpha \simeq \lim_{m\rightarrow\infty}\frac{F_{m-1}}{F_{m}}$, where $F_{m}$ is defined recursively through the relation $F_{m+1}=F_{m}+F_{m-1}$, with $F_{0}=F_{1}=1$ \cite{wang2016phase,cai2021localization}. To see the PBC in numerics, we choose the number of lattice site $L$ to be one number of the Fibonacci series \cite{wang2016phase,deng2019one,wang2021unconventional}.

The energy spectrum as the function of disorder strength $V$ under PBC is shown in Fig. \ref{figure:1}. It is shown that there is a threshold of $V$, i.e., $V_{c}$. When $V$ is below that critical value $V_{c}$, the spectrum remains gapped, at $V=V_c$ the gap of spectrum vanishes. When further increasing $V$, the system becomes fully gapped again. Accordingly, the critical disorder strength can be determined by the gap-closing point when varying $V$. We further find that the critical $V_{c}$ will be shifted with varying the asymmetric hopping amplitude $\gamma$. As shown in Figs. \ref{figure:1}(a) and \ref{figure:1}(b), for $\gamma=1t$ and $1.4t$, the corresponding gap-closing points $V_{c}$ are $3.4t$ and $3.7t$, respectively. Compared to the Hermitian case with $\gamma=0$, the gap-closing point $3.0t$, determined by the relation $V_{c}=2(t+\Delta)$ \cite{cai2013topological}. To illustrate this effect more clearly, we show the $V_{c}$ as a function of $\gamma$ in Fig. \ref{figure:1}(c). It is shown that $V_{c}$ will be enlarged when increasing $\gamma$.

The stability and existence of MZMs are inherently associated with the above changes in spectrum. To further demonstrate the effect of the non-local disorder on MZMs, we introduce the Majorana operators $\kappa_{i}^{A} =\hat{c}_{i}^{\dagger }+ \hat{c}_{i}$ and $\kappa_{i}^{B} = i(\hat{c}_{i}^{\dagger }-\hat{c}_{i})$. These operators adhere to the relations $(\kappa_i^\alpha)^{\dagger} = \kappa_i^\alpha$ and $\{{\kappa _{i}^{\alpha },\kappa _{i}^{\beta }\}}=2\delta _{ij}\delta_{\alpha \beta }$, where $\alpha$ and $\beta$ refer to $A$ or $B$. The quasiparticle operators are reformulated in terms of Majorana operators as
\begin{eqnarray}
\eta _{n}^{\dag } = \frac{1}{2}\sum_{i=1}^{L}[ \varphi _{n, i}
\kappa_i^A  - i \psi _{n,i} \kappa_i^B ],
\end{eqnarray}
where $\varphi _{n, i}=(u_{n, i}+ v_{n, i})$ and $ \psi _{n,i}=(u_{n, i} - v_{n, i})$.
\begin{figure}[tbp]	
		\centering
        \includegraphics[width=0.75\textwidth]{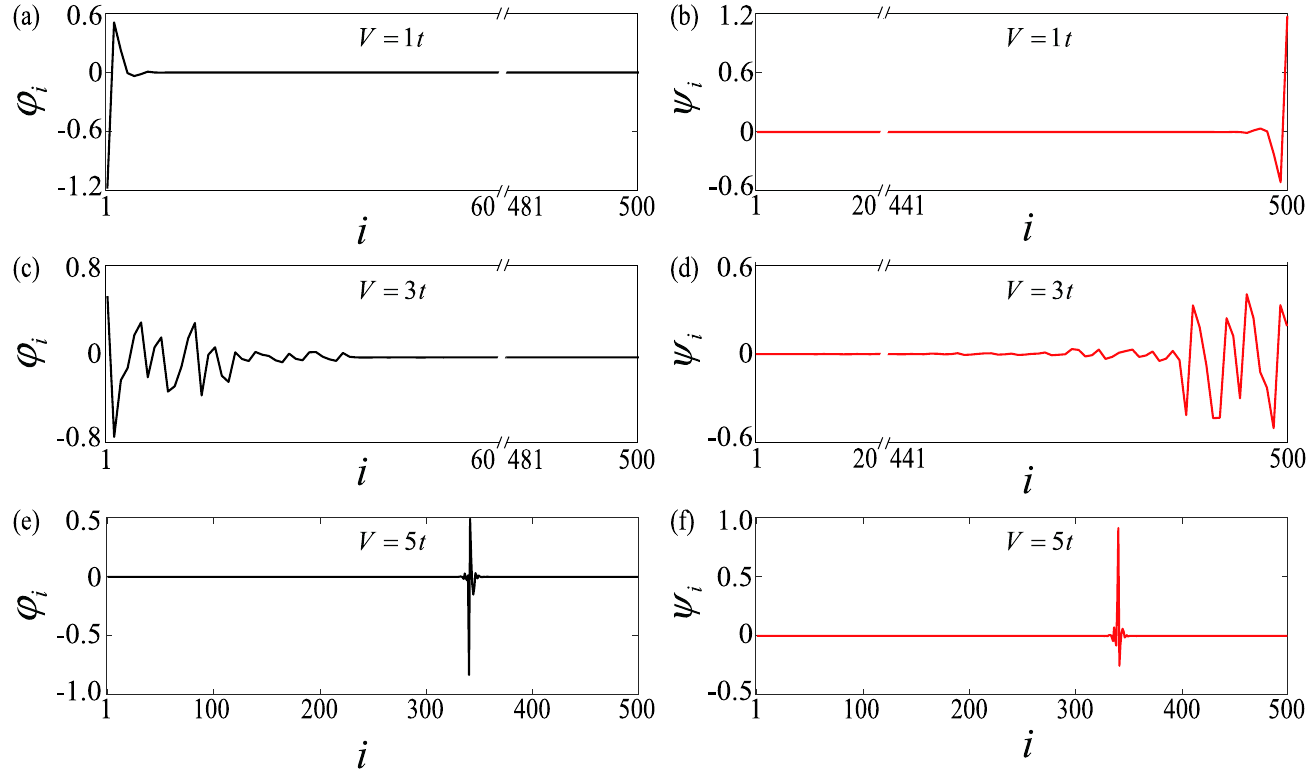}	
		\caption{Spatial distributions of $\varphi_i$ (left panel) and $\psi_i$ (right panel) for the lowest excitation solution for different $V$. From top to bottom, the disorder strength $V$ increases from $1t$ (a, b) to $3t$ (c, d), and finally to $5t$ (e, f). With increasing $V$, the MZMs are changed from being localized at the two ends of the chain to totally all overlapped, become normal fermions. Here, we choose $\alpha =(\sqrt{5}-1)/2$, $\gamma/t=1$, $\Delta/t=0.5$ and $L=500$.}
		\label{figure:mfs}		
	\end{figure}
Figure \ref{figure:mfs} illustrates the distributions of $\varphi_i$ and $\psi_i$ for the lowest excitation solution under open boundary condition (OBC). As shown in Figs. \ref{figure:mfs}(a) and \ref{figure:mfs}(b), for $V<V_c$, two MFs corresponding to the zero-mode solution in the gapped SC phase are spatially separated and localized at each end of the chain. As $V$ increasing to $V_c$, the energy gap decreases, and MFs at each end of the system will be overlapped. This leads to a rapid decay of the edge modes away from the left (right) edge, as depicted in Figs. \ref{figure:mfs}(c) and \ref{figure:mfs}(d). Conversely, when $V>V_c$, the corresponding quasiparticle becomes a localized normal fermion that cannot be decomposed into two independent MFs, as shown in Figs. \ref{figure:mfs}(e) and \ref{figure:mfs}(f). This signifies the phase transition from a topologically nontrivial SC phase to a trivial phase. Remarkably, this phase transition is suppressed by increasing non-Hermiticity. Through examining the energy spectrum and wave function of the lowest energy state via varying $\gamma$, it can be concluded that non-Hermiticity significantly enhances the robustness of MZMs against non-local disorder, since $V_c$ will increase when enlarging $\gamma$.
\begin{figure}[thbp]
\begin{center}
\includegraphics[width=0.6\textwidth]{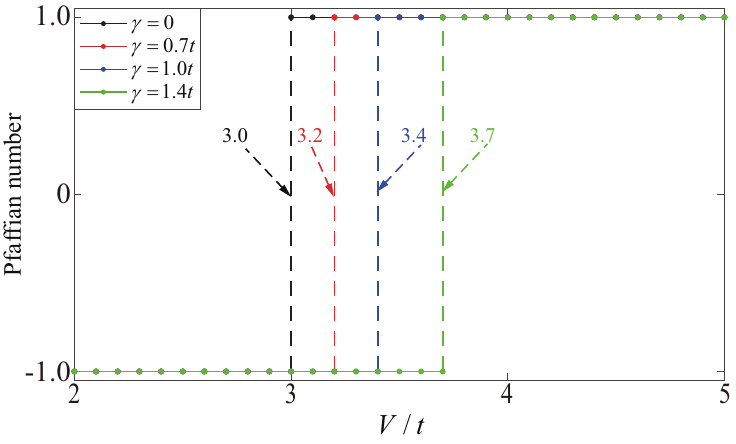}
\end{center}
\caption{Pfaffian number as the function of $V$ under OBC with $\alpha =(\sqrt{5}-1)/2$, $\Delta/t=0.5$ and $L=500$. The topologically nontrivial phase is characterized by $M = -1$, whereas the trivial phase corresponds to $M = 1$.}
\label{figure:topin}%
\end{figure}
Such existence of MZMs actually arises from the nontrivial topological properties of the bulk superconductor, which can be characterized by a topological invariant, i.e., Pfaffian number. We introduce the Majorana basis $(\kappa _{1}^{A},\kappa _{1}^{B},\kappa _{2}^{A},\kappa _{2}^{B}, \dots, \kappa_{L}^{A},\kappa _{L}^{B})^{T}$ as $(\kappa _{1},\kappa _{2},\kappa _{3},\kappa_{4}, \dots,\kappa _{2L-1},\kappa _{2L})^{T}$ and the Hamiltonian (\ref{Eq.Hamiltonian}) can be reexpressed as $\mathbf{\tilde{H}}=\frac{i}{4}\sum\limits_{l,m}^{2L}Q_{lm}^{^{\prime}}\kappa_{l}\kappa_{m}$, where $Q$ is a $2L \times 2L$ matrix. The Pfaffian number is defined as $M=\mathrm{sgn}(\mathrm{Pf}(Q))$, where $\mathrm{Pf}(Q)=\frac{1}{2^LL!}\sum_{\tau\in S_{2L}}\mathrm{sgn}(\tau)Q_{\tau(1),\tau(2)}Q_{\tau(3),\tau(4)} \dots Q_{\tau(2L-1),\tau(2L)}$, and $S_{2L}$ is the set of permutations on $2L$ elements and $\mathrm{sgn}(\tau)$ is the sign of permutation \cite{kitaev2001unpaired,cai2013topological}. As shown in Fig. \ref{figure:topin}, when $V < V_c$, the Pfaffian number evaluates to $M = -1$ for all $\gamma$, indicating a topologically nontrivial phase. Conversely, when $V > V_c$, the Pfaffian number is changed to $M = 1$. A topological phase transition to the trivial phase occurs as $V$ increases. Furthermore, as the asymmetric hopping $\gamma$ increases, the topological phase transition point shifts to higher values, being consistent with the $V_c$ determined by the gap-closing condition. This observation substantiates our conclusion that non-Hermiticity improves the stability of the immunity of the topological phase.

\section{\label{sec3}Analytical derivation of the critical $V_{c}$ }
\begin{figure}[tbp]	
		\centering
        \includegraphics[width=0.6\textwidth]{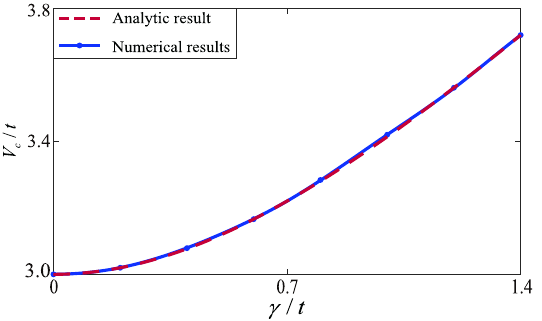}	
		\caption{Analytical result(red dashed line) and numerical results(blue solid line) for $V_{c}$ curve with respect to the changes of $\gamma$. All parameters are the same as in Fig. \ref{figure:1}.}
		\label{figure:naa}		
	\end{figure}

In the following, we will analytically determine the critical $V_{c}$ in the large-$L$ limit by using the gap-closing condition. The Hamiltonian matrix can be written as $\mathcal{H}=\begin{pmatrix}
A & B \\
-B & -A^{T}
\end{pmatrix}$,
where $A$ and $B$ are $L \times L$ matrices and can be expressed as
\begin{equation}
A=%
\begin{pmatrix}
V_{1} & -(t+\gamma /2) & ... & -(t-\gamma /2) \\
-(t-\gamma /2) & V_{2} & -(t+\gamma /2) & ... \\
... & -(t-\gamma /2) & ... & -(t+\gamma /2) \\
-(t+\gamma /2) & ... & -(t-\gamma /2) & V_{L}%
\end{pmatrix}%
,B=%
\begin{pmatrix}
0 & -\Delta  & ... & \Delta  \\
\Delta  & 0 & -\Delta  & ... \\
... & \Delta  & ... & -\Delta  \\
-\Delta  & ... & \Delta  & 0%
\end{pmatrix},%
\end{equation}%
where $V_{i}=V\cos (2\pi \alpha i )$. Additionally, matrix $A$ can be decomposed into a symmetric matrix $A_{t}$ and an antisymmetric matrix $A_{g}$, such that $A=A_{t}+A_{g}$, with
\begin{equation}
A_{t}=%
\begin{pmatrix}
V_{1} & -t & ... & -t \\
-t & V_{2} & -t & ... \\
... & -t & ... & -t \\
-t & ... & -t & V_{L}
\end{pmatrix}%
,A_{g}=%
\begin{pmatrix}
0 & -\gamma /2 & ... & \gamma /2 \\
\gamma /2 & 0 & -\gamma /2 & ... \\
... & \gamma /2 & ... & -\gamma /2 \\
-\gamma /2 & ... & \gamma /2 & 0%
\end{pmatrix}%
.
\end{equation}%
Therefore, the excitation spectrum $E_{n}$ can be determined by solving the following equation,
\begin{equation}
\det(\mathcal{H}-E_{n})=
\begin{vmatrix}
(A_{g}+A_{t})-E_{n} & B \\
-B & (A_{g}-A_{t})-E_{n}%
\end{vmatrix}%
=0.  \label{Eq.deterAE}
\end{equation}
The solution of Eq. (\ref{Eq.deterAE}) at the critical $V_c$ satisfies the relation $E_{n}=0$, and the critical $V_c$ can thus be determined by solving $\det (\mathcal{H})=0$. The determinant of $\mathcal{H}$ can be further transformed as
\begin{equation}
\begin{vmatrix}
A_{g}+A_{t} & B \\
-B & A_{g}-A_{t}%
\end{vmatrix}%
\overset{c_{1}+kc_{2}}{=}%
\begin{vmatrix}
A_{t} & B \\
-\frac{\gamma }{2\Delta }(A_{g}-A_{t})-B & A_{g}-A_{t}%
\end{vmatrix}%
\overset{r_{2}+kr_{1}}{=}%
\begin{vmatrix}
A_{t} & B \\
\frac{\gamma }{2\Delta }A_{g}-B & -A_{t}%
\end{vmatrix}%
\overset{mc_{2}}{\underset{r_{2}/m}{=}}%
\begin{vmatrix}
A_{t} & C \\
-C & -A_{t}%
\end{vmatrix}%
,  \label{7}
\end{equation}
where $k=-\frac{\gamma }{2\Delta }$, $m=\frac{\sqrt{\Delta ^{2}+\gamma ^{2}/4%
}}{\Delta }$, $C=%
\begin{pmatrix}
0 & -\sqrt{\Delta ^{2}+\gamma ^{2}/4} & ... & \sqrt{\Delta ^{2}+\gamma ^{2}/4%
} \\
\sqrt{\Delta ^{2}+\gamma ^{2}/4} & 0 & -\sqrt{\Delta ^{2}+\gamma ^{2}/4} &
... \\
... & \sqrt{\Delta ^{2}+\gamma ^{2}/4} & ... & -\sqrt{\Delta ^{2}+\gamma
^{2}/4} \\
-\sqrt{\Delta ^{2}+\gamma ^{2}/4} & ... & \sqrt{\Delta ^{2}+\gamma ^{2}/4} &
0%
\end{pmatrix}%
$, $r_i$ and $c_i$ $(i=1,2)$ denote the rows and columns of the determinant, respectively.
Through defining $\mathcal{N}=%
\begin{pmatrix}
A_{t} & C \\
-C & -A_{t}%
\end{pmatrix}%
$, it is shown that $\det (\mathcal{H})=\det (\mathcal{N})$. The determinant of $\mathcal{N}$ can subsequently be transformed as
\begin{equation}
\begin{vmatrix}
A_{t} & C \\
-C & -A_{t}%
\end{vmatrix}%
\overset{r_{1}-r_{2}}{=}%
\begin{vmatrix}
A_{t}+C & C+A_{t} \\
-C & -A_{t}%
\end{vmatrix}%
=|C+A_{t}|%
\begin{vmatrix}
I & I \\
-C & -A_{t}%
\end{vmatrix}%
\overset{r_{2}+Cr_{1}}{=}|C+A_{t}|%
\begin{vmatrix}
I & I \\
O & C-A_{t}%
\end{vmatrix}%
=|C+A_{t}||C-A_{t}|.  \label{8}
\end{equation}
Consequently, solving $\det (\mathcal{H})=\det (\mathcal{N})=0$ can be converted to $\det[(C+A_{t})(C-A_{t})]=0$. By applying the relation $\det (C-A_{t})=\det (C-A_{t})^{T}=\det (C+A_{t})$, the critical value $V_c$ can be determined through
\begin{equation}
\det (C-A_{t})=0.\label{Eq.deterAC}
\end{equation}
After long but straightforward algebra (see the Appendix for details), the analytical solution for $V_{c}$ for our model in Eq. (\ref{Eq.Hamiltonian}) is derived as
\begin{equation}
\left\vert V_{c}\right\vert =2(\left\vert t\right\vert +\sqrt{\Delta
^{2}+\gamma ^{2}/4}).  \label{Eq.analytic_eq}
\end{equation}
As illustrated in Fig. \ref{figure:naa}, the analytical result (red dashed line) for infinite $L$ aligns closely with the numerical results (blue solid line), being consistent with Fig. \ref{figure:1}(c). Equation (\ref{Eq.analytic_eq}) demonstrates that the critical $V_{c}$ increases as $\gamma$ increases, reflecting the enhanced robustness of the immunity of the topological SC phase against non-local disorder.

\section{\label{sec4}Summary}

In summary, we examine the effect of the non-Hermiticity on the immunity of the topological SC phase in the 1D Kitaev model against non-local disorder. Through introducing the asymmetric hopping amplitudes, it can enlarge the critical disorder strength, beyond that the MZMs disappeared. This finding suggests that non-Hermiticity can improve the robustness of MZMs against non-local disorder. By doing both numerical and analytical calculations, we show that these two approaches are consistent with each other well. Our findings not only advance the understanding of the interplay between non-Hermiticity and topology, but also are potentially useful for disorder-resistant quantum computing platforms based on topological superconductors.

\section*{Acknowledgements}
This work is supported by the National Key R$\&$D Program of China (2021YFA1401700), NSFC (Grants No. 12474267, 12074305), the Fundamental Research Funds for the Central Universities (Grant No. xtr052023002), the Shaanxi Fundamental Science Research Project for Mathematics and Physics (Grant No. 23JSZ003), Shanghai Municipal Science and Technology Major Project (Grant No.2019SHZDZX01) and the Xiaomi Young Scholar Program. We also thank the HPC platform of Xi'an Jiaotong University, where our numerical calculations was performed.

\renewcommand{\thesection}{A-\arabic{section}}
\setcounter{section}{0}  
\renewcommand{\theequation}{A\arabic{equation}}
\setcounter{equation}{0}  
\renewcommand{\thefigure}{A\arabic{figure}}
\setcounter{figure}{0}  

\indent
\section*{\label{Appendix}Appendix. Details for the derivation of the critical $V_{c}$}
\appendix
In this Appendix, we derive the condition of the critical $V_{c}$ of the model
Hamiltonian in Eq. (\ref{Eq.Hamiltonian}). The detailed derivation to obtain Eq. (\ref{Eq.analytic_eq}). Starting from Eq. (\ref{Eq.deterAC}), the determinant $\mathrm{det}(C-A_{t})$ can be calculated:
\begin{eqnarray}
&&\mathrm{det}(C-A_{t})  \notag \\
&=&-(t-\sqrt{\Delta ^{2}+\gamma ^{2}/4})^{L}-(t+\sqrt{\Delta ^{2}+\gamma
^{2}/4})^{L}+V^{L}\prod_{i=1}^{L}\mathrm{cos}(2\pi \alpha i)  \notag \\
&&+V^{L-2}(\Delta ^{2}+\gamma ^{2}/4-t^{2})\sum\limits_{j=1}^{L}\prod
_{\substack{ i=1 \\ i\neq j,j+1}}^{L}\mathrm{cos}(2\pi \alpha
i)+V^{L-4}(\Delta ^{2}+\gamma
^{2}/4-t^{2})^{2}\sum\limits_{j_{1}=1}^{L}\sum\limits_{j_{2}=j_{1}+2}^{L}%
\prod_{\substack{ i=1 \\ i\neq j_{1},j_{1}+1, \\ j_{2},j_{2}+1}}^{L}\mathrm{%
cos}(2\pi \alpha i)  \notag \\
&&+\cdots +V^{L-2n}(\Delta ^{2}+\gamma ^{2}/4-t^{2})^{n}\sum\limits
_{\substack{ \{j_{s}=j_{s-1}+2, \\ s=1,\cdots ,n\}}}^{L}\prod_{\substack{ i=1
\\ i\neq j_{s},j_{s}+1 \\ (s=1,\cdots ,n)}}^{L}\mathrm{cos}(2\pi \alpha
i)+\cdots .  \label{sumdeter}
\end{eqnarray}%

When the irrational number $\alpha= (\sqrt{5}-1)/2$ is approximated by $\alpha \simeq \lim_{m\rightarrow\infty}\frac{F_{m-1}}{F_{m}}$, where $F_{m}$ is the Fibonacci number and $m$ is an integer, Without loss of generality, we can assume that $\alpha =p/q$ with $q\rightarrow \infty $ to approach a given irrational number, where $p$ and $q$ are co-prime integers. For a finite-sized system, distinguishing an irrational $\alpha $ and its rational approximation $p/q$ becomes physically unfeasible when $q\geq L$ \cite{cai2013topological,cai2021localization}. In the following we evaluate the determinant of the $L\times L$ matrix by taking $\alpha =p/q$ with $L=q$, where $p$ is co-prime to $L$ (for the inverse golden ratio $L=F_{m}$).

Under the condition of $\alpha =p/L$, we find that Eq. (\ref{sumdeter}) can be simplified as the summation terms including $\mathrm{cos}(2\pi \alpha i)$, which are vanished for each $m$ except $m=0$ and $L/2$ (if $L$ is even). For examples,  when $L=3$, the term including $\mathrm{cos}(2\pi \alpha i)$ except $m=0$ is $\sum\limits_{i=1}^{3}\cos (\frac{2\pi pi}{3})=0$; when $L=6$, the terms including $\mathrm{cos}(2\pi \alpha i)$ except $m=0$ and $3$ are $\sum \limits_{i=1}^{6}\cos(\frac{2\pi pi}{6})\cos[\frac{2\pi p(i+1)}{6}]\cos[\frac{2\pi p(i+2)}{6}]\cos[\frac{2\pi p(i+3)}{6}]=0$ and $\sum\limits_{i=1}^{6}\cos(\frac{2\pi pi}{6}) \cos[\frac{2\pi p(i+1)}{6}]+\sum\limits_{i=1}^{3}\cos(\frac{2\pi pi}{6})\cos[\frac{2\pi p(i+3)}{6}]=0$. For larger $L$, it can be shown that all the terms including $\mathrm{cos}(2\pi \alpha i)$ vanish except $m=0$ and $L/2$, only leaving $V^{L}\prod_{i=1}^{L}\cos(2\pi \alpha i)$ and $(\Delta^{2}+\gamma ^{2}/4-t^{2})^{L/2}$. Therefore, we can obtain $\mathrm{det}(C-A_{t})$ as
\begin{equation}
\det (C-A_{t})=\left\{
\begin{array}{l}
\prod_{i=1}^{L}V\cos (2\pi \alpha i)-(t-\sqrt{\Delta ^{2}+\gamma ^{2}/4}%
)^{L}-(t+\sqrt{\Delta ^{2}+\gamma ^{2}/4})^{L},\ \text{for odd }L, \\
\prod_{i=1}^{L}V\cos (2\pi \alpha i)-(t-\sqrt{\Delta ^{2}+\gamma ^{2}/4}%
)^{L}-(t+\sqrt{\Delta ^{2}+\gamma ^{2}/4})^{L}+(\Delta ^{2}+\gamma
^{2}/4-t^{2})^{L/2},\ \text{for even }L.%
\end{array}%
\right.   \label{resoddeven}
\end{equation}
Based on the above discussion, the critical $V_{c}$ determined by Eq. (\ref{Eq.deterAC}) satisfies the following relation
\begin{equation}
\left\{
\begin{array}{l}
\prod_{i=1}^{L}\frac{V_{c}}{\sqrt{\Delta ^{2}+\gamma ^{2}/4}+t}\mathrm{\cos }%
(2\pi \alpha i)-(\frac{t-\sqrt{\Delta ^{2}+\gamma ^{2}/4}}{t+\sqrt{\Delta
^{2}+\gamma ^{2}/4}})^{L}-1=0,\ ~\text{for odd }L, \\
\prod_{i=1}^{L}\frac{V_{c}}{\sqrt{\Delta ^{2}+\gamma ^{2}/4}+t}\mathrm{\cos }%
(2\pi \alpha i)-(\frac{t-\sqrt{\Delta ^{2}+\gamma ^{2}/4}}{t+\sqrt{\Delta
^{2}+\gamma ^{2}/4}})^{L}+(\frac{\sqrt{\Delta ^{2}+\gamma ^{2}/4}-t}{\sqrt{%
\Delta ^{2}+\gamma ^{2}/4}+t})^{L/2}-1=0,\ ~\text{for even }L.%
\end{array}%
\right.\label{VcforOE}
\end{equation}%
Here, we assume that $t>0$, and in the thermodynamic limit  $L\rightarrow \infty $, Eq. (\ref{VcforOE}) will be reduced to
\begin{equation}
\prod_{i=1}^{L}\mathrm{\cos }(2\pi \alpha i)=\left( \frac{\sqrt{\Delta
^{2}+\gamma ^{2}/4}+t}{V_{c}}\right) ^{L} \label{VcforOEred}
\end{equation}%
Taking the logarithm of both sides of Eq. (\ref{VcforOEred}), Eq. (\ref{VcforOEred}) can thus be expressed as
\begin{equation}
\frac{1}{L}\sum_{i=1}^{L}\ln \cos (2\pi \alpha i)=\ln \left( \frac{\sqrt{%
\Delta ^{2}+\gamma ^{2}/4}+t}{V_{c}}\right) +\frac{i2\pi m}{L}. \label{Eq.lnequi}
\end{equation}%
Meanwhile, the summation on the left-hand side of the above equation can be replaced by an integral in the limit $L\rightarrow \infty $, that is%
\begin{equation}
\frac{1}{L}\sum_{i=1}^{L}\ln \cos (2\pi \alpha i)\rightarrow \int_{0}^{1}\ln
\cos (2\pi \alpha Lx)dx=\frac{1}{2\pi \alpha L}\int_{0}^{2\pi \alpha L}\ln
\mathrm{\cos }(x)dx=-\frac{1}{2\pi \alpha L}\mathcal{L}(2\pi \alpha L),\label{intesum}
\end{equation}%
where $\mathcal{L}(z)$ is the Lobachevskiy's function \cite{gradshteyn2014table} defined as
\begin{equation}
\mathcal{L}(z)=-\int_{0}^{z}\ln \cos (x)dx=z\ln 2-\frac{1}{2}%
\sum_{k=1}^{\infty }(-1)^{k-1}\frac{\sin (2kz)}{k^{2}}. \label{Lobfun}
\end{equation}%
Through substituting Eq. (\ref{intesum}) into Eq. (\ref{Lobfun}), in the limit $L\rightarrow \infty $,  we obtain the relation
\begin{equation}
\lim_{L\rightarrow \infty }\frac{1}{L}\sum_{i=1}^{L}\ln \cos(2\pi\alpha i)=-\ln 2. \label{Eq.Lobach}
\end{equation}%
Combining Eq. (\ref{Eq.lnequi}) and Eq. (\ref{Eq.Lobach}), we can get that the critical $V_{c}$ should be satisfied the following relation
\begin{equation}
\left\vert V_{c}\right\vert =2(\sqrt{\Delta ^{2}+\gamma ^{2}/4}+t).
\end{equation}%
The above derivation is not limited to the case of $t>0$, for general cases, similar derivation can be directly followed and $V_{c}$ is given by
\begin{equation}
\left\vert V_{c}\right\vert =2(\sqrt{\Delta ^{2}+\gamma ^{2}/4}+\left\vert t\right\vert ),
\end{equation}
which gives the phase transition point from topologically nontrivial SC phase to trivial phase.
\bibliographystyle{elsarticle-num}
\bibliography{enhancerefer}

\end{document}